\documentclass{article}

\usepackage{arxiv}

\usepackage[utf8]{inputenc} 
\usepackage[T1]{fontenc}    
\usepackage{graphicx}
\usepackage{subfigure}
\usepackage{courier}
\usepackage{xcolor}
\usepackage{booktabs}
\usepackage{tabularx}
\usepackage{amsmath,amssymb,amsfonts,flushend}
\usepackage{algorithm}
\usepackage[colorlinks=true, urlcolor=blue, citecolor=blue, linkcolor=blue]{hyperref}

\title{Thread and Data Mapping in Software Transactional Memory: An Overview}

\author{ \href{https://orcid.org/0000-0001-6125-8994}{\includegraphics[scale=0.08]{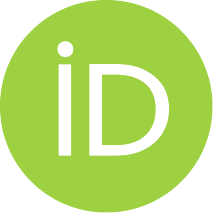}\hspace{1mm}Douglas Pereira Pasqualin} \\
	Computer Science Graduate Program (PPGC) \\ Universidade Federal de Pelotas \\ 
	Rio Grande do Sul, 96075-630, Brazil \\
	\texttt{dp.pasqualin@inf.ufpel.edu.br} \\
	\And
	\href{https://orcid.org/0000-0002-9064-7806}{\includegraphics[scale=0.08]{orcid.pdf}\hspace{1mm}Matthias Diener} \\
	University of Illinois Urbana-Champaign \\ Illinois, 61801, USA \\
	\texttt{mdiener@illinois.edu} \\
	\And
\href{https://orcid.org/0000-0002-6790-5184}{\includegraphics[scale=0.08]{orcid.pdf}\hspace{1mm}Andr\'e Rauber Du Bois} \\
	Computer Science Graduate Program (PPGC) \\ Universidade Federal de Pelotas \\ 
	Rio Grande do Sul, 96075-630, Brazil \\
	\texttt{dubois@inf.ufpel.edu.br} \\
	\And
\href{https://orcid.org/0000-0003-2382-2755}{\includegraphics[scale=0.08]{orcid.pdf}\hspace{1mm}Maur\'icio Lima Pilla} \\
	Computer Science Graduate Program (PPGC) \\ Universidade Federal de Pelotas \\ 
	Rio Grande do Sul, 96075-630, Brazil \\
	\texttt{pilla@inf.ufpel.edu.br}	
}

\date{}

\renewcommand{\shorttitle}{Thread and Data Mapping in STM}

\hypersetup{
pdftitle={Thread and Data Mapping in Software Transactional Memory: An Overview},
pdfsubject={preprint},
pdfauthor={Douglas Pereira Pasqualin, Matthias Diener, André Rauber Du Bois, Maurício Lima Pilla },
pdfkeywords={Software Transactional Memory, Thread Mapping, Data Mapping, Communication},
}

\begin{document}
\maketitle

\begin{abstract}
In current microarchitectures, due to the complex memory hierarchies and different latencies on memory accesses, thread and data mapping are important issues to improve application performance. Software transactional memory (STM) is an abstraction used for thread synchronization, replacing the use of locks in parallel programming. Regarding thread and data mapping, STM presents new challenges and mapping opportunities, since (1) STM can use different conflict detection and resolution strategies, making the behavior of the application less predictable and; (2) the STM runtime has precise information about shared data and the intensity with each thread accesses them. These unique characteristics provide many opportunities for low-overhead, but precise  statistics to guide mapping strategies for STM applications. The main objective of this paper is to survey the existing work about thread and data mapping that uses solely information gathered from the STM runtime to guide thread and data mapping decisions. We also discuss future research directions within this research area.
\end{abstract}


\section{Introduction}

Since the year 2000, multicore processors have become the main configuration of new processor developments. This decision was taken due to microarchitectural limitations, higher power consumption, and heat dissipation involved in improving the performance of a single CPU core~\cite{Trono:2015}. 
Since then, the number of cores in a single chip has been growing every year. Besides, servers normally have multiple multicore processors, where each socket is connected directly to a \emph{local} memory module~\cite{Gaud:2015}. These architectures are called NUMA (Non-Uniform Memory Access) systems and are becoming dominant in servers~\cite{Calciu:2017}. Accessing data that is mapped to a memory module that belongs to a different processor, that is, in a \emph{remote node}, implies a higher latency, making the access time non-uniform, that is, depending on the location of the data. 

Due to multicore processors being widely available, parallel programming is becoming even more important. A crucial issue that arises in parallel programming is thread synchronization; locks are still the most used abstraction for this purpose. However, locks often make source code difficult to read and debug, leading to problems such as deadlocks~\cite{Herlihy:2020,Anthes:2014}. An alternative abstraction to replace mutual-exclusion locks in parallel programming is \emph{Transactional Memory} (TM)~\cite{Grahn:2010}, in which critical sections are accessed using transactions similar to the ones available in databases. The TM runtime is responsible for ensuring a consistent execution, for example, without deadlocks and race conditions. A transaction that has executed without conflicts can commit, that is, update the memory with the new values. If a conflict is detected, an abort is executed and a transaction is reinitialized until a commit is possible. TMs can be implemented in hardware (HTM), software (STM), or both (hybrid).
HTM has the advantage of a lower overhead compared to STM since they do not need instrumentation to track transactional operations~\cite{Didona:2016}. However, HTM has resource limitations. For example, when the footprint of a transaction exceeds the L1 cache capacity, it is aborted~\cite{Diegues:2014:2}. In that case, software alternatives are necessary to guarantee progress. This paper focuses on STM implementations since all prior proposals found in the literature are based on STM.

Another important issue in multicore processors is the mapping of threads and data. Due to complex memory hierarchies and different latencies in memory accesses on these processors, thread and data placement policies that improve the use of memory controllers and data locality are important to achieve good performance. Although thread and data mapping policies have been widely studied for general purpose parallel applications~\cite{Diener:2016Sur} there is a lack of research in the STM domain. More specifically about thread mapping, STM adds new challenges such as different kinds of conflict detection and resolution. Hence, the best thread mapping depends on the STM configuration~\cite{Castro:2011}. In addition, the STM runtime has precise information about memory areas that are shared between threads, their respective memory addresses, and the intensity with which they are accessed by each thread, providing interesting mapping opportunities~\cite{Pasqualin:2021}. In that case, a thread mapping based only on STM accesses will have a lower overhead than other proposals that focus on general applications, since it is not necessary to keep track of all memory accesses of the applications to determine the mapping.

In this paper, we describe and survey proposals that focus on using the information provided by the STM runtime to guide thread and data mapping. Previous surveys on thread and data mapping focused only on general applications~\cite{Diener:2016Sur}. On the other hand, previous surveys on STM focused on scheduling techniques in order to improve the performance~\cite{Hendler:2015,Sanzo:2017}. 
To the best of our knowledge, there are no prior surveys of thread and data mapping in the context of STM.

We found that the STM runtime provides accurate information about the sharing behavior of the application and is thus able to provide sufficient information for an improved thread mapping. For data mapping, our survey showed that it might be unfeasible to rely only on the information provided by the STM runtime to perform an efficient data mapping since the STM runtime usually has access to only a fraction of the entire memory accessed by the application, limiting the accuracy of a data mapping algorithm.

The remainder of this paper is organized as follows. The next section presents the main concepts of STM as well as thread and data mapping and briefly describes other techniques used to improve the performance of TM. Section~\ref{sect:survey} surveys and discusses the related work on thread and data mapping in the context of STM. We also present a table comparing and classifying all discussed works. In Section~\ref{sect:directions}, we propose new research directions to fill the gaps missing in the related work. Finally, Section~\ref{sect:conclusion} concludes the paper.

\section{Background}\label{sect:background}
\subsection{Software Transactional Memory}\label{sect:STM} 

Transactional memory (TM) is an abstraction to synchronize accesses to shared variables. Instead of using locks, the programmer only needs to enclose the critical section in an atomic block, which will be executed as a transaction. The concept of transactions was borrowed from Databases. The first TM implementation purely on software (STM) was proposed by Shavit and Touitou~\cite{Shavit:1995}. The execution of a transaction needs to be \emph{atomic}. Atomicity requires that a transaction is executed as a whole or it needs to appear as it was never executed~\cite{Grahn:2010}. A transaction \emph{commits} if executed without conflicts, hence all operations and results are made visible to the rest of the system. If conflicts are detected, a transaction \emph{aborts}, that is, all operations are discarded, and the transaction needs to restart until a commit is possible. This idea is associated with another important property called \emph{isolation}: all memory updates of a running transaction can not be visible to other transactions before a commit.

Although the main purpose of STM is to provide a simple interface to manage access to shared data, its implementation is not trivial. Many different design options are available such as transaction granularity, version management, conflict detection, and resolution. The \emph{granularity} is the dimension used for conflict detection, that is, the level used for keeping track of memory locations. For STM, the most used options are \textbf{word} or \textbf{object}. \emph{Version management} manages the writes of concurrent transactions to memory locations. It can be \textbf{eager}, where data is modified directly in memory and an \emph{undo-log} is used to restore old values in case of aborts; or \textbf{lazy}, where data is updated in a \emph{redo-log}. During the commit, the log is used to set new values to memory. \emph{Conflict Detection} uses the same nomenclature as version management. In \textbf{eager} conflict detection, conflicts are verified on each memory location accessed. To access a value, a transaction needs to acquire its ownership. In \textbf{lazy} conflict detection, the ownership will be acquired during the commit phase. In case of conflicts, choosing which transaction needs to be aborted is a responsibility of the \emph{Contention Manager} (CM). There are many CMs proposed in the literature with different purposes. The most simple action that a CM can do is to abort the transaction that detected the conflict.

Regarding STM implementation, \texttt{TinySTM}~\cite{TinySTM2} is one of the most used STM libraries and is considered a state-of-art STM implementation~\cite{Chen:2020,Sanzo:2020}. \texttt{TinySTM} uses word granularity and has configurable version management and CMs. With respect to benchmarks, \texttt{STAMP} (\textit{Stanford Transactional Applications for Multi-Processing})~\cite{STAMP} still is the most used STM benchmark suite~\cite{Pasqualin:Bench}. This suite is composed of 8 applications with realistic characteristics, covering a wide range of transactional behavior, representing several application domains.

\subsection{Thread Mapping}\label{sect:threadMapping}

In thread mapping, threads are associated to cores, improving the cache usage and machine interconnections. The strategy utilized depends on the application and the underlying architecture. For instance, in applications where a group of threads accesses the same shared data, these threads can be mapped to cores that are close to each other in the underlying architecture. If the application has a disjoint data access pattern, where each group of threads accesses different groups of data, mapping them on different cores processors can improve performance, since more cache will be available to each thread. The default scheduler used by the Linux kernel, called \emph{Completely Fair Scheduler} (CFS)~\cite{Wong:2008} mainly focuses on load balancing. If the behavior of an application is known in advance and it does not change during execution, a static thread mapping can be used. For instance, \figurename~\ref{fig:MappingStrategy} shows different static thread mapping strategies with distinct objectives:

\begin{figure}[!bt]
	\centering
	\subfigure[Compact.]{
		\label{fig:MappingStrategyCompact}
		\includegraphics[width=0.5\textwidth,trim=0 0 0 17,clip]{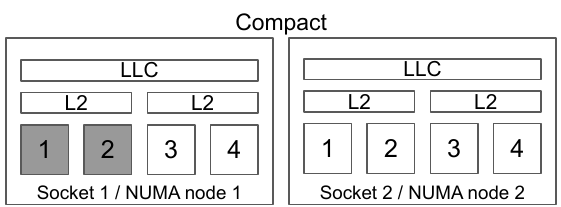}
	}%
	\\
	\subfigure[Scatter.]{
		\label{fig:MappingStrategyScatter}
		\includegraphics[width=0.5\textwidth,trim=0 0 0 17,clip]{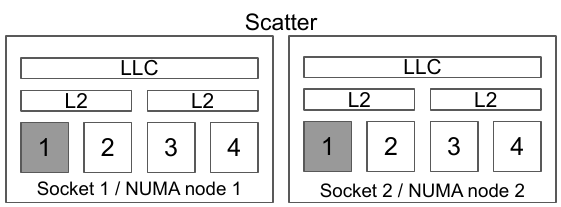}
	}%
	\\
	\subfigure[Round-Robin.]{
		\label{fig:MappingStrategyRR}
		\includegraphics[width=0.5\textwidth,trim=0 0 0 17,clip]{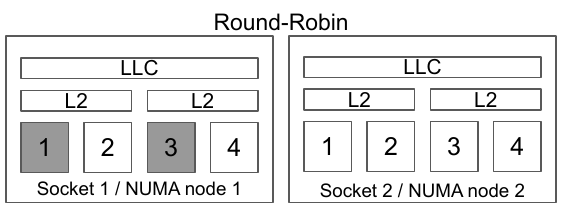}
	}%
	\caption{Thread mapping strategies.}
	\label{fig:MappingStrategy}
\end{figure}

\begin{itemize}
	\item \textbf{Compact} places threads on sibling cores that share all cache levels, thus potentially reducing the data access latency if neighboring threads communicate often (\figurename~\ref{fig:MappingStrategyCompact}).
	\item \textbf{Scatter} distributes threads across different processors, avoiding cache sharing, thus, reducing memory contention (\figurename~\ref{fig:MappingStrategyScatter}).
	\item \textbf{Round-Robin} is a mix between compact and scatter, where only the last level of cache (LLC) is shared (\figurename~\ref{fig:MappingStrategyRR}). It is worth noting that if the processor architecture employs lower cache levels as private to each core, for instance, the L2, then the resulting mapping will be exactly the same as Compact.
\end{itemize}

\begin{figure}[!ht]
	\centering
	\subfigure[Numbered matrix.]{
		\includegraphics[width=0.25\textwidth]{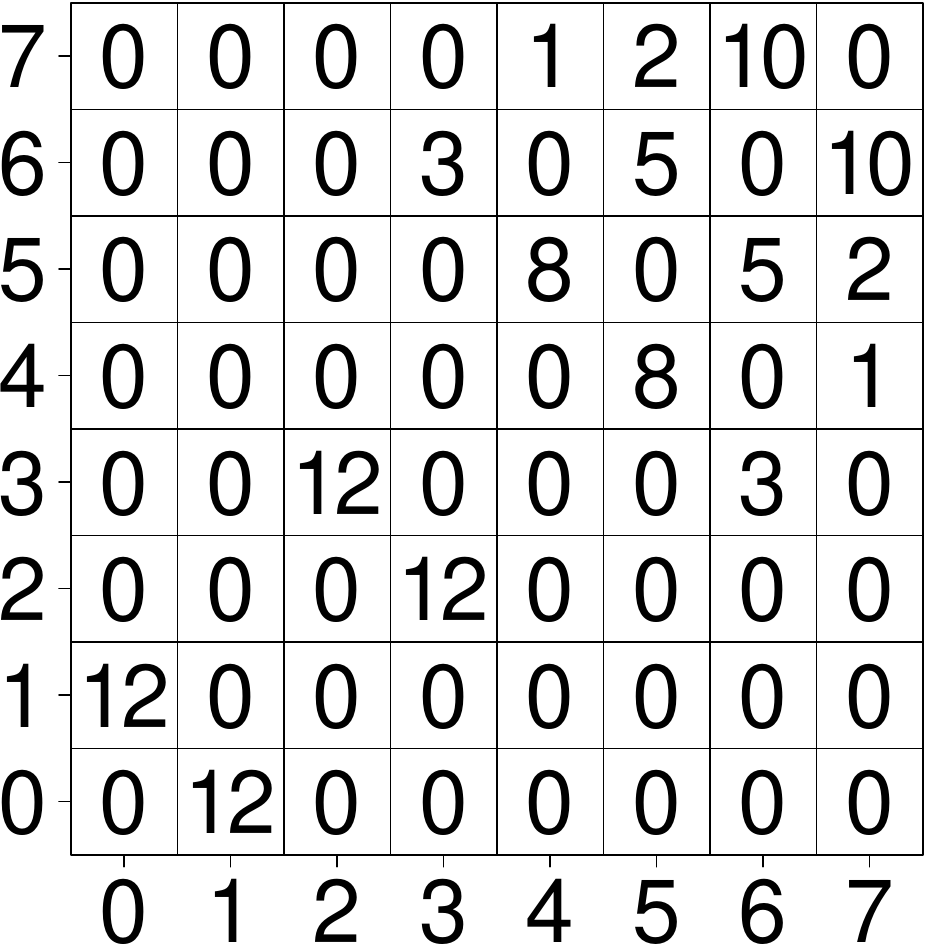}
		\label{fig:comMatrNum}
	}
\hspace*{1cm}
	\subfigure[Heatmap.]{
		\label{fig:comMatrGraph}
		\includegraphics[width=0.25\textwidth]{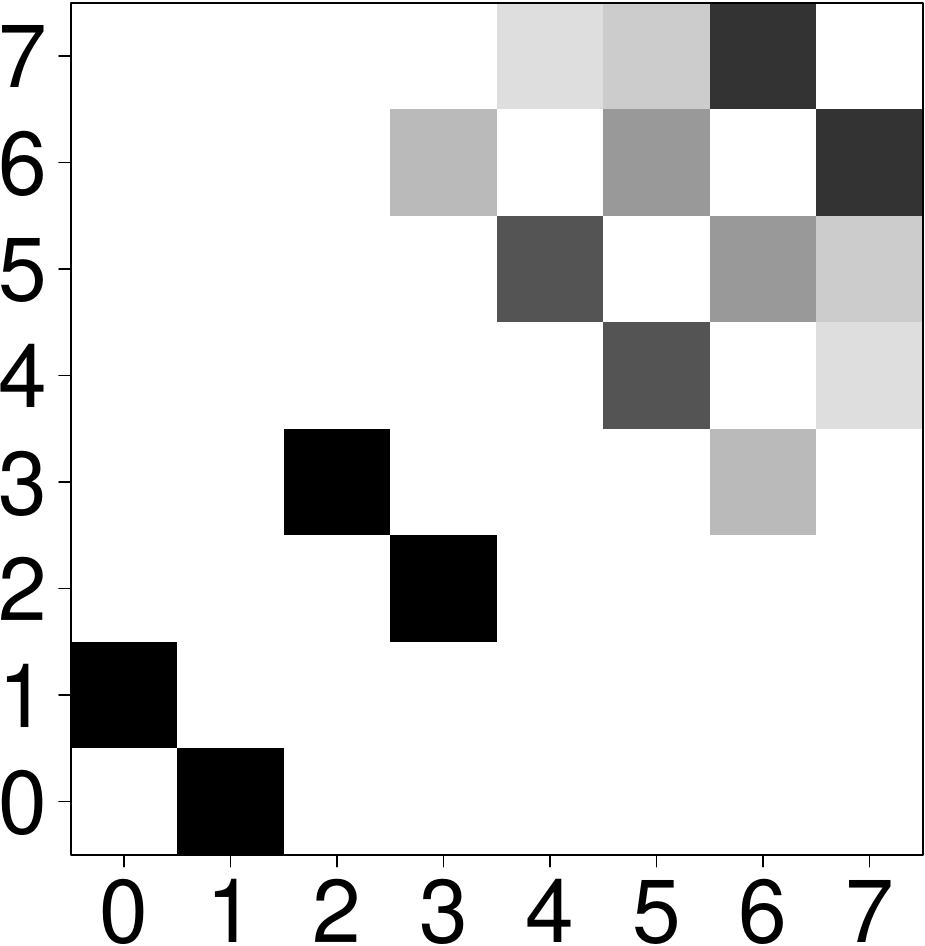}
	}
	\caption{Examples of communication matrices.}
	\label{fig:comMatrExample}
\end{figure}

Regarding locality of the memory access, some techniques are used to keep track of the sharing behavior of the application in order to map threads to cores based on the memory access behavior. Thread mapping based on the memory access behavior of applications is called sharing-aware\footnote{This research field is also known as affinity-based~\cite{Diener:2016Sur} or topology-aware~\cite{Jeannot:2013} mapping.} thread mapping~\cite{Cruz:2018}. There are many techniques in literature to perform an efficient sharing-aware thread mapping~\cite{Diener:2016Sur}. One of the main challenges of this area in shared memory architectures is to detect which threads are accessing each memory address. An affinity measure is necessary to be able to quantify the groups of threads that have more affinity. A communication or sharing matrix is the most common measure to determine the affinity between threads~\cite{Bordage:2018, Mazaheri:2018}. Each cell in the matrix represents the amount of communication between pairs of threads. \figurename~\ref{fig:comMatrExample} shows examples of communication matrices, where axes are thread IDs. 
In \figurename~\ref{fig:comMatrGraph}, the matrix is represented graphically, where darker cells indicate more communication between pairs of threads.

The communication matrix can be used as an input to task-mapping algorithms. These algorithms use the hierarchical topology of the machine and the communication matrix to calculate an optimized mapping of threads to cores. Many tools are available to calculate a thread mapping based on a communication matrix, for instance, \texttt{Scotch}~\cite{Pellegrini:1994}, \texttt{TreeMatch}~\cite{TreeMatch}, \texttt{EagerMap}~\cite{Cruz:2019}, \texttt{ChoiceMap}~\cite{Soomro:2018}, and \texttt{TopoMatch}~\cite{TopoMatch}.

\subsection{Data Mapping}\label{sect:dataMapping}

In data mapping, memory pages are associated to NUMA nodes, optimizing the usage of memory controllers and interconnections. Similar to thread mapping, the strategy utilized depends on the application. The most common strategies are load balance or locality. In \emph{locality}, memory pages are mapped to the same NUMA node where the core that most accesses them is located. The Linux kernel implements data mapping strategies in the form of memory allocation policies. The default policy is called \emph{first-touch}~\cite{Gaud:2015} where memory is allocated in the NUMA node where the first access to the memory page is performed. Another data mapping policy available is \emph{interleave}, which focuses on balance, allocating pages in a round-robin way on the NUMA nodes~\cite{Lameter:2013}.

The default Linux kernel already has routines to improve memory locality of NUMA nodes. It keeps track of page faults, moving a page automatically to the node that most accessed it. This mechanism is called \emph{NUMA balancing}~\cite{NumaB:2020}.

\subsection{Other Techniques To Improve TM Performance}\label{sect:others}

This section briefly describes other techniques used to improve the performance of TM. The main objective of most techniques is to improve performance by avoiding conflicts, that is, reducing the number of aborts.

The first technique, \emph{Transactional scheduling}, acts in a \emph{proactive} way, using heuristics to prevent conflicts and to decide \emph{when} and \emph{where} a transaction should be executed \cite{Shrink}. When the use of scheduling techniques in STM first appeared, the main idea was to avoid conflicts, and the solution, in general, is to serialize a conflicting transaction~\cite{Hendler:2015}.

Another technique, \emph{Concurrency control}~\cite{Ansari:2014}, limits the number of concurrent threads running transactions. The idea is that an excessive number of threads can diminish performance in a high-contention environment, mainly due to a higher number of aborts. Di Sanzo~\cite{Sanzo:2017} classifies concurrency control mechanisms as \emph{Thread scheduling}. Transactional scheduling and thread scheduling were surveyed in~\cite{Hendler:2015, Sanzo:2017}.

Regarding \emph{Hardware Transactional Memory (HTM)}, when the hardware aborts a transaction, for instance, when the footprint of a transaction exceeds the L1 cache capacity or in context switches, the hardware will abort the transaction. In such a case, software alternatives (fallback paths) are necessary to guarantee progress. These techniques are surveyed in~\cite{Wu:2020}.
 
\section{Thread and Data Mapping in STM}\label{sect:survey}

\begin{table}[tb]
	\centering
	\caption{Comparison of the proposed works to deal with thread and data mapping in STM applications.}
	\label{tab:comparison}
	\footnotesize

\begin{tabular}{@{}llcclclc@{}}
	\toprule
	Work                                     & Year & \multicolumn{1}{l}{\begin{tabular}[c]{@{}l@{}}Thread\\ mapping\end{tabular}} & \multicolumn{1}{l}{\begin{tabular}[c]{@{}l@{}}Data\\ mapping\end{tabular}} & \begin{tabular}[c]{@{}l@{}}Mapping\\ offline or \\ online\end{tabular} & \multicolumn{1}{l}{\begin{tabular}[c]{@{}l@{}}Needs prior\\ knowledge of \\ the application?\end{tabular}} & \begin{tabular}[c]{@{}l@{}}Data used to define \\ the mapping\end{tabular}                                                                       & \multicolumn{1}{l}{\begin{tabular}[c]{@{}l@{}}Changes \\ to the \\ application\end{tabular}} \\ \midrule
	Castro et al.~\cite{Castro:2011}         & 2011 & \checkmark                                                                              &                                                                            & Offline                                                                & Yes                                                                                                           & \begin{tabular}[c]{@{}l@{}}Abort ratio, LLC miss ratio,\\ transaction time ratio and\\ ML algorithm\end{tabular}                                 & No                                                                                           \\[0.5cm]
	Castro et al.~\cite{Castro:2012}         & 2012 & \checkmark                                                                              &                                                                            & Both                                                                   & Yes                                                                                                           & \begin{tabular}[c]{@{}l@{}}Abort ratio, LLC miss ratio,\\ transaction time ratio and\\ ML algorithm\end{tabular}                                 & No                                                                                           \\[0.5cm]
	Castro et al.~\cite{Castro:2014}         & 2014 & \checkmark                                                                              &                                                                            & Both                                                                   & No                                                                                                            & \begin{tabular}[c]{@{}l@{}}Abort ratio, LLC miss ratio,\\ transaction time ratio and\\ ML algorithm\end{tabular}                                 & No                                                                                           \\ \midrule
	Góes et al.~\cite{Goes:2012}             & 2012 & \checkmark                                                                              &                                                                            & Online                                                                 & Yes                                                                                                           & Sharing pattern (worklist)                                                                                                                       & Yes                                                                                          \\[0.15cm]
	Góes et al.~\cite{Goes:2014}             & 2014 & \checkmark                                                                              & \checkmark                                                                            & Both                                                                   & Yes                                                                                                           & Sharing pattern (worklist)                                                                                                                       & Yes                                                                                          \\ \midrule
	Chan et al.~\cite{Chan:2015}             & 2015 & \checkmark                                                                              &                                                                            & Online                                                                 & No                                                                                                            & \begin{tabular}[c]{@{}l@{}}Sharing pattern \\ (based on aborts)\end{tabular}                                                                     & No                                                                                           \\ \midrule
	Zhou et al.~\cite{Zhou:2016}             & 2016 & \checkmark                                                                              &                                                                            & Online                                                                 & No                                                                                                            & \begin{tabular}[c]{@{}l@{}}Throughput \\ (between all mappings)\end{tabular}                                                                     & No                                                                                           \\[0.3cm]
	Zhou et al.~\cite{Zhou:2018}             & 2018 & \checkmark                                                                              &                                                                            & Online                                                                 & No                                                                                                            & Throughput                                                                                                                                       & No                                                                                           \\ \midrule
	Pasqualin et al.~\cite{Pasqualin:2020}   & 2020 & \checkmark                                                                              &                                                                            & Offline                                                                & Yes                                                                                                           & \begin{tabular}[c]{@{}l@{}}Sharing pattern (based on\\ reads and writes)\end{tabular}                                                            & No                                                                                           \\[0.3cm]
	Pasqualin et al.~\cite{Pasqualin:2020:2} & 2020 & \checkmark                                                                              &                                                                            & Online                                                                 & No                                                                                                            & \begin{tabular}[c]{@{}l@{}}Sharing pattern (based on\\ reads and writes), \\ abort ratio and \\ number of accessed memory\\ addresses\end{tabular} & No                                                                                           \\[0.8cm]
	Pasqualin et al.~\cite{Pasqualin:2021}   & 2022 &                                                                              & \checkmark                                                                          & Online                                                                 & No                                                                                                            & \begin{tabular}[c]{@{}l@{}}Sharing pattern (based on\\ reads and writes)\end{tabular}                                                            & No                                                                                           \\ \bottomrule
\end{tabular}
\end{table}

This section surveys the related work on thread and data mapping in the context of software transactional memory. The works described in this section are summarized in \tablename~\ref{tab:comparison}. When describing the works in the following sections, we opted to group the works by authors since the majority of the authors proposed more than one approach on this subject. 

\subsection{Castro et al.}\label{sect:castro}

To the best of our knowledge, the first work to use transactional information to guide thread mapping was proposed by Castro et al.~\cite{Castro:2011}. The motivation is that STM adds new challenges to thread mapping, such as different kinds of conflict detection and resolution. Hence, the best thread mapping depends on the STM configuration. To illustrate this scenario, the authors showed an experiment using all applications from the \texttt{STAMP}~\cite{STAMP} benchmark using three different state-of-art STM libraries,  \texttt{TinySTM}~\cite{TinySTM2}, \texttt{SwissSTM}~\cite{SwissTM} and \texttt{TL2}~\cite{TL2} and the three thread mappings as shown in \figurename~\ref{fig:MappingStrategy}. The best mapping varies according to the STM used. This is due to the different strategies used for each STM to deal with conflicts.

The authors proposed to feed a machine learning (ML) algorithm to discover the best thread mapping for each application. The ML algorithm used was the \textit{Iterative Dichotomiser 3} (\texttt{ID3})~\cite{ID3} which is based on decision trees. For each kind of conflict resolution and contention management strategy, they decided to collect the abort ratio (aborts divided by the total of transactions), LLC miss ratio (cache misses divided by the total accesses) and, transaction time ratio (transaction time / total execution time). After that, a second step normalizes the data and removes duplicates. This processed information is used as an input to  \texttt{ID3} for the training step. The final result is a decision tree called \emph{predictor}, that is able to predict the best thread mapping according to the input data. It is worth noting that the machine used to collect the information influences the predictor, that is, it is machine-dependent. Analyzing the generated predictor, they concluded that the abort ratio and LLC miss ratio are strongly related to the mapping predicted. For instance, when the abort ratio is low, the preferred mapping is scatter, as the application accesses few shared data.

Finally, the generated predictor was implemented inside the \texttt{TinySTM}~\cite{TinySTM2} library and, using the \texttt{STAMP}~\cite{STAMP} benchmark, it was possible to improve the execution time of applications, even when compared to an oracle that gives the best case.

The main drawback of the proposed approach is that thread mapping is chosen by the predictor only once during the execution of the application. However, for applications whose behavior changes during runtime, a more dynamic approach to adapt thread mapping for each new phase might be necessary. Hence, Castro et al.~\cite{Castro:2012} extended their previous work, creating an approach that changes the thread mapping dynamically during runtime. The basic approach is the same as the previous work: use \texttt{ID3} to generate a predictor being able to predict the best thread mapping according to the input parameters. A profiling interval and an interval between profiles are used to collect the necessary information for the predictor. Both intervals are based on a predefined amount of commits. The interval begins short and doubles each time that the thread mapping is not changed. When the mapping changes, the interval is reset to the initial value. It is worth noting that, to avoid high overhead, only one thread is used to be the profiler thread. Since, according to the authors, the \texttt{STAMP} benchmarks do not present dynamic behavior during runtime, they used the synthetic \texttt{EigenBench}~\cite{EigenBench} benchmark to test the proposed mechanism. Using this benchmark, they created 56 synthetic applications, each one presenting 3 distinct execution phases. Overall the proposed dynamic mechanism switched correctly to the best thread mapping on the majority of applications.

In \cite{Castro:2014},  the authors tested a new ML algorithm, called \texttt{Apriori}~\cite{Apriori}. Another contribution is a new dynamic thread mapping technique where it is not necessary to have any prior knowledge of the applications. Different from the  ML approaches used in their previous work, the proposed technique does not need preliminary runs of applications to collect information to guide mapping. As described in~\cite{Castro:2011}, the abort ratio has a strong relationship with the best thread mapping. Hence, the idea is to use this single metric to choose an optimal thread mapping dynamically while the application is running. The application starts with the default mapping chosen by the default Linux scheduler. After that, a profiler collects information about the abort ratio during $n$ committed transactions. When the profiling phase ends, according to the abort ratio, a thread mapping is chosen between compact, scatter, and round-robin. Then, this new thread mapping is active until the next profiling phase ends. This strategy was called \texttt{Conflict}. Beyond that, another strategy called \texttt{Test-and-map} was proposed. In this strategy, the profiling phase is divided into three parts. On each part one thread mapping is activated and the execution time is measured. When the profile phase ends, the mapping that achieved the shortest execution time is selected to be active until the next profile time ends.

All proposed strategies, that is, \texttt{ID3} and \texttt{Apriori} that are based on ML and \texttt{Conflict} and \texttt{Test-and-map} that are based on a single metric, were tested using the \texttt{STAMP}~\cite{STAMP} and the 56 synthetic applications constructed with \texttt{EigenBench}~\cite{EigenBench}. All strategies showed performance gains when compared to the default Linux scheduler. As expected, the ML approaches showed higher gains, since they use more information to decide the mapping.

\subsection{Góes et al.}\label{sect:goes}

Goés et al.~\cite{Goes:2012} used the concept of skeleton or pattern-based programming, where a parallel program has a well-defined communication and computation pattern during execution. More specifically, they focused on a \textit{worklist} pattern, where each \textit{work unit} to be processed is dynamically managed by a \textit{worklist}, that is, a collection of work unit instances. This pattern was chosen by observing that the majority of the \texttt{STAMP}~\cite{STAMP} applications have this pattern. 

They proposed the \texttt{OpenSkel} framework in the form of a runtime system library to be used for transactional worklist applications. This library provides an API~(\textit{Application Programming Interface}) to deal with transactional worklists, applying performance optimizations and autotuning during runtime. These tunings are possible due to pattern-based programming used, as applications present a well-defined computational pattern. Thus, it is possible to know in advance what an application will do next. The main primitives provided by \texttt{OpenSkel} are to allocate, run, and free a worklist. It should be noted that a transactional application needs to be designed or modified to use the proposed \texttt{OpenSkel} framework. The worklist is represented internally by \texttt{OpenSkel} using a stack data structure, allowing to improve data locality if consecutive work units access common shared data. 

As the STM does not manage thread scheduling and mapping, the OS is responsible for this task. Since the default Linux scheduler tries to perform load balancing, and migrate threads during runtime, the authors opted for a fixed \textit{scatter} thread mapping strategy statically when the application starts. Hence, since the thread mapping was set manually, the OS will not be allowed to migrate threads during runtime, allowing it to have more predictable performance. Besides, according to the authors, scatter is the mapping more similar to the default Linux strategy of load balancing.

\texttt{OpenSkel} proposes optimizations to improve performance. One specific technique is the use of helper threads (HTs) that do not change the state of the application. Their main objective is to prefetch data and bring them to the caches. HTs can be used if there are idle cores in the system, specifically, idle cores that share cache levels with cores running transactions.

For the tests, they chose five applications from the \texttt{STAMP}~\cite{STAMP} benchmark suite that present a worklist pattern: intruder, kmeans, labyrinth, vacation, and yada. These applications were modified to use the \texttt{OpenSkel} framework. The underlying STM system used to manage transactional operations was \texttt{TinySTM}~\cite{TinySTM2}. The results indicate that, in general, the proposed framework improves the performance of applications, including when compared to an oracle approach.

In \cite{Goes:2014}, the \texttt{OpenSkel} framework was extended, adding a new module responsible for dealing with memory affinity. The main focus is to improve the performance of STM applications on NUMA architectures, by using page allocation policies. The HTs used for prefetching data together with page allocation policies were integrated into a mechanism called \texttt{SkelAff}, a submodule of \texttt{OpenSkel}, responsible to improve the memory affinity. Since the framework knows the next work unit that needs to be executed, \texttt{SkelAff} can trigger an HT to prefetch data that will be required soon in advance. In addition, it can allocate memory pages close to a specific thread if the mechanism knows in advance that there are work-units in the queue to be executed that are memory-related to work-units being executed. The memory page policies can be \textit{bind} or \textit{cyclic}. The \textit{bind} allocation reduces the access latency by binding data used by a thread on a single NUMA node. On the other hand, \textit{cyclic} distributes memory pages using a round-robin strategy, focusing on load balance. By using information about the work-units, \texttt{OpenSkel} chooses the best memory page allocation policy and triggers HTs when necessary. 

For the experiments, the focus was to verify the efficiency of the proposed \texttt{SkelAff} mechanism. They chose four applications from \texttt{STAMP}~\cite{STAMP} benchmark that present the worklist pattern: intruder, kmeans, vacation, and yada (labyrinth was not used in these experiments) and two different NUMA machines. Results indicate that the proposed framework improves the performance of applications and that \textit{cyclic} page allocations deliver more performance gains.

\subsection{Chan et al.}\label{sect:chan}

In Chan et al.~\cite{Chan:2015} two mechanisms were proposed. The first one is a dynamic concurrency control mechanism, used to limit the maximum number of concurrent threads executing transactions. The idea is that an excessive number of threads can hurt performance in a high contention environment, mainly due to a higher number of aborts. Thus, it is better to have a lower number of threads executing transactions than a high number of aborts.  The second proposed mechanism keeps track of transactional conflicts to detect which threads are accessing the same shared data. Hence, based on this information, an \emph{affinity-aware thread migration} is proposed. The main objective is to improve the cache usage of shared data managed by the STM runtime, consequently improving the execution time through enhanced cache locality. Also, they focus on multiprocessor systems. Since the concurrency control mechanism (Sect.~\ref{sect:others}) is out of the scope of this paper, we will focus on the affinity-aware thread migration approach.


When a conflict is detected, before aborting, the mechanism verifies the identifier (id) of the conflicting thread. Hence, this relationship is updated in a matrix. The value stored in each matrix position is derived from the formula of \emph{contention intensity}, initially proposed by Yoo and Lee~\cite{Yoo:2008}. The derived formula was called \emph{pairwise contention intensity}~(PCI). The value stored in the $i,j$ position of the matrix means how likely thread $i$ is to be obstructed by thread $j$. The matrix can be seen as a directed adjacency matrix, where nodes represent threads and edges represent weights are the PCI between a pair of threads. Using a graph partitioning algorithm, the objective is to reduce the sum of edges that span across different cores processors. To avoid overhead, only one pair of threads is migrated between processors each time that the mechanism is triggered.

During a commit, the committing transaction, for instance, transaction $i$, updates its conflict intensity with all other threads (transactions), that is, the PCI must be computed for all other running threads and updated in the matrix. The PCI used during the commit phase is: $C_{ij} \leftarrow C_{ij} \times 0.9$. On the other hand, when a transaction aborts, the PCI against the conflicting transaction is updated in the matrix, using the formula $C_{ij} \leftarrow C_{ij} \times 0.9 + (1 - 0.9)$. Periodically, a daemon thread applies a graph partition algorithm in the matrix to choose the pair of threads that should be migrated. It should be noted that the algorithm to calculate the migration policy was designed to work on dual-processor systems only.

Both mechanisms, concurrency control and affinity-aware thread migration, were implemented inside the \texttt{TinySTM}~\cite{TinySTM2} library. Threads were pinned to cores using the \mbox{\texttt{pthread\_setaffinity\_np}} function. For the experiments, the authors used the \texttt{STAMP}~\cite{STAMP} benchmarks. In addition, the proposed mechanism was compared with the transactional schedulers ATS~\cite{Yoo:2008} and Shrink~\cite{Shrink}. In \texttt{STAMP}, the affinity-aware thread migration does not improve performance. According to the authors, \texttt{STAMP} does not present any specific thread correlation. However, to demonstrate the efficiency of the proposed technique, the authors created a synthetic dual Red-Black tree application, where it was possible to control the thread correlation, thus making it possible to improve the performance with the proposed thread migration mechanism.

\subsection{Zhou et al.}\label{sect:zhou}

Similar to Chan et al. (Sect.~\ref{sect:chan}), Zhou et al.~\cite{Zhou:2016} also proposed a concurrency control mechanism to dynamically adjust the maximum number of threads allowed to run transactions concurrently. As the number of active threads changes, the thread mapping is changed as well. Two key pieces of information are used by the authors in their proposed mechanism: the commit ratio (CR) calculated by the total of commits divided by the total of commits plus aborts and; throughput, that is, the total of commits in a unit of time. The motivation to use both metrics are twofold: (1) throughput is influenced by the number of threads, that is, a low throughput can be a consequence of a low number of threads running and;(2) CR is influenced by throughput, that is, a low CR can present a high throughput if there is a high number of threads running transactions.

The mechanism starts with a probabilistic model to calculate the predicted optimum number of threads allowed to run concurrently. During a profiling phase, the throughput is measured and compared to the previous one collected before changing the number of threads. If the current throughput is lower than the previous, the number of threads is switched back to the previous value. A second step is related to thread mapping. After deciding the thread number, if the CR changed (in a predefined range) and the thread number is less than half of the maximum core number, a second profile phase begins to decide the thread mapping. According to the authors, thread mapping has little impact on the performance when the number of cores is closer to the maximum cores of the machine, where all mappings perform similarly. For this reason, thread mapping is only calculated when the number of concurrent threads running is less than half of the cores of the machine. During the thread mapping profiling phase, four mappings are tested: Linux, scatter, compact, and Round-robin. In the end, the thread mapping that presented the higher throughput is set. After that, the mechanism restarts. An important aspect is that the information used by the profile is collected on all running threads, instead of, as in other works, based on a single thread. Hence, synchronization costs are added to gather the required profile information.

For the experiments, 6 applications from \texttt{STAMP}~\cite{STAMP} (bayes and kmeans were not used) and one synthetic from \texttt{EigenBench}~\cite{EigenBench} were used. According to the authors, \texttt{STAMP} applications present similar behavior between transactions, that is, overall, each transaction executes the same transactional code. Hence, they are not suitable to evaluate online thread mapping strategies. For this specific purpose, \texttt{EigenBench} was used. They conclude that due to the overhead, the proposed mechanism can benefit only applications that present dynamic behavior variation during execution. Finally, the authors discovered that thread mapping is influenced by the CR. In most cases, when the CR was low, compact was chosen, whereas when the CR was high, scatter was preferred.

In \cite{Zhou:2018}, together with the probabilistic model, the authors also proposed a \emph{simple model} that searches for the near-optimum thread number. In this model, the thread number is incremented or decremented by one, according to the measured throughput in each profile phase. Regarding thread mapping, adjustments were made in order to reduce overhead. In their previous work, all mappings were tested during the profile phase, measuring the throughput and setting the mapping at the end. In the new approach, the thread mapping profile starts with Linux and then moves to the round-robin mapping. The compact mapping is tested only if round-robin throughput was better than Linux. Finally, scatter is tested only if the throughput of compact was worse than round-robin. When scatter is the best mapping, the overhead will be the same as the previous approach since all mappings will be tested.

The experiments and conclusions were similar to their previous paper. Nevertheless, they also concluded that the proposed mechanism does not scale on NUMA architectures, since the proposed mechanism does not take into consideration memory location and has excessive thread migrations to test the best thread mapping.

\subsection{Pasqualin et al.}\label{sect:pasqualin}

Pasqualin et al.~\cite{Pasqualin:2020} used sharing-aware thread mapping~(Sect.~\ref{sect:threadMapping}) in the context of STM, to map threads to cores considering their memory access behavior. Contrary to previous sharing-aware mapping proposals that rely on memory traces of the entire application, they argue that tracking only STM operations to determine the sharing behavior has lower overhead and better accuracy because only memory accesses that are in fact shared between threads are traced. The main intuition used is that STM has precise information about shared variables and has native access to all information needed to characterize the sharing behavior, that is, accessed memory addresses and the intensity with which each thread accesses them. 

One of the main challenges of sharing-aware thread mapping in shared memory architectures is to detect which threads are accessing each memory address. However, word-based STM has native access to these pieces of information since they need to keep track and save versions of shared reads and writes to data. The proposed mechanism is triggered on each transactional read or write operation. An auxiliary hash table maps memory addresses to the last 2 threads that accessed them. When at least 2 distinct threads access the same memory address, a communication event between them is updated in a communication or shared matrix. The communication matrix generated by the proposed mechanism is sent to \texttt{EagerMap}~\cite{Cruz:2019}, a task mapping algorithm for sharing-aware mapping, to generate the optimized thread mapping.

The proposed mechanism to collect the communication matrix was implemented inside the STM library \texttt{TinySTM}~\cite{TinySTM2}. It is necessary to run each application individually in order to extract the communication matrix. After that, \texttt{EagerMap} generates the optimized thread to core mapping, and the application is re-executed pinning threads to cores using the function \mbox{\texttt{pthread\_setaffinity\_np}}. 

For the experiments, the authors chose \texttt{STAMP}~\cite{STAMP} and two synthetic applications, \emph{Hashmap} and \emph{Red-black tree}. The comparison of the proposed mechanism was made against compact, scatter, and round-robin mappings. Also, applications were executed in a NUMA machine using Opteron processors. Not all tested applications were suitable for the proposed sharing-aware mapping,  including reducing the number of aborts. Finally, it is demonstrated that the overhead to collect the communication matrix is much lower than using other tools that trace all memory addresses, such as \texttt{numalize}~\cite{numalize}.

In~\cite{Pasqualin:2020:2}, the proposed mechanism was modified to collect the communication matrix and perform thread mapping during the execution of an application. The new mechanism was called \texttt{STMap}. To reduce the overhead of the mechanism, the authors sampled the accessed memory addresses to fill the communication matrix. They have made experiments to determine the best \emph{sampling interval} (SI) to have a low overhead but a high accuracy of the collected communication matrix. Thus, a SI of 100 was selected. Besides, it was necessary to define the \emph{mapping interval} (MI), that is, when the thread mapping should be calculated and performed. Based on experiments and observations, they chose an MI of between 50,000 and 100,000 addresses accessed (total, not only sampled) by the main thread. Regarding calculating the thread mapping, in \texttt{STMap}, \texttt{EagerMap}~\cite{Cruz:2019} was replaced by \texttt{TopoMatch}~\cite{TopoMatch}. 

Since \texttt{STMap} works during runtime, the authors tried to create a heuristic to identify the applications that are not suitable for sharing-aware thread mapping, thus, disabling \texttt{STMap} for these applications, that is, not applying the new thread mapping during the execution. The heuristic is based on the number of distinct memory addresses accessed by the application and the abort and commit ratios. The new thread mapping should be calculated if; when the MI is triggered the application accessed less than 10,000 distinct memory addresses. If more than 10,000 addresses were accessed, the heuristic verifies if the abort ratio is greater than the commit ratio. If true, the thread mapping should be calculated as well. 

The methodology of the experiments was similar to the previous work. Nevertheless, they included an Xeon machine in the experiments and the previous static mechanism to the comparison. The proposed heuristic to identify which applications are suitable for sharing-aware thread mapping worked correctly for the majority of applications, improving execution time when compared to the static mechanism. Hence, taking into consideration the improved execution time over all benchmarks, \texttt{STMap} was the best mapping.

Finally, in~\cite{Pasqualin:2021} \texttt{STMap} was extended to include sharing-aware data mapping. However, the authors conclude that, contrary to thread mapping where only taking into consideration STM access is sufficient to have a global vision of the sharing behavior, for data mapping this observation is not enough. For data mapping, it is necessary to have a global vision of memory pages accessed to be able to perform an optimized data mapping, not only the memory pages accessed by the sharing data protected by STM runtime.

\subsection{Discussion}

We will start by discussing data mapping. As can be observed in \tablename~\ref{tab:comparison}, just two works attempted to deal with data mapping in STM. The first one proposed by Góes et al.~\cite{Goes:2014} provided a mechanism to choose whether the data allocated inside the STM runtime should be allocated in the same NUMA node as the thread that needs it or prioritizing a memory balance. Hence, it is based on a static (offline) mapping. However, as discussed by Pasqualin et al.~\cite{Pasqualin:2021} it could be infeasible to perform an efficient online data mapping by only having accesses to the memory pages accessed by the STM runtime since it represents only a fraction of the entire memory accessed by the application. As the opposite, for thread mapping, the information provided by the STM runtime is exactly the one needed to perform an efficient thread mapping. One of the main objectives of thread mapping is to improve cache usage by mapping cores that access the same shared data in a way that they can share different cache levels.

Regarding thread mapping, the information about aborts, such as amount and ratio compared to commits, can be a useful metric to define thread mapping. The intuition used is that if the aborts are high, then the application is accessing a great quantity of shared data. Hence, it can be interesting to use a mapping strategy that prioritizes locality. In fact, many works used abort information to determine the thread mapping.

There are works that instead of relying solely on the number of aborts to identify a global shared pattern, try to identify which specific threads are accessing the same shared data. Chan et al.~\cite{Chan:2015} tried to identify the relationship between threads when they abort. However, their mechanism has a high overhead since it is triggered on each abort and the affinity measured should be recalculated for all threads each time. Pasqualin et al.~\cite{Pasqualin:2020:2} used a similar idea, but tried to improve the accuracy of the mechanism by tracking transactional reads and writes, instead of only aborts. The mechanism has a lower overhead since the affinity measure is a counter that is updated once for each access. Additionally, transactional reads and writes are only sampled. 

A different strategy was proposed by Zhou et al.~\cite{Zhou:2016,Zhou:2018} to identify the best thread mapping. When the number of concurrent threads executing transactions is changed, a new thread mapping is set to match the new number of threads. Their mechanism measures the throughput of each thread mapping and chooses the one that presents a higher throughput at the end of the test phase. Although the proposed mechanism works correctly, it incurs high overhead because of the thread migrations triggered to test all mappings. As pointed out by the authors, this mechanism does not scale well on NUMA machines due to the excessive thread migrations.

Some mechanisms require previous knowledge of the applications. Using this knowledge, it is possible to perform an offline thread mapping at the beginning of the execution of the STM application. Although the main disadvantage of these offline mechanisms is the need for a preliminary run to gather the necessary information about the application, during the execution of the application, they present low overhead and can be more accurate and efficient compared to online mechanisms~\cite{Agullo:2016}. Another related disadvantage is that an offline mechanism can be inefficient if the application changes the sharing behavior during the execution. In contrast, an online mechanism does not need a preliminary run to gather information about applications and can adapt if the sharing behavior suddenly changes during runtime. However, depending on the chosen strategy, the online mechanism can generate an infeasible overhead to be used during runtime.

Finally, an important characteristic of the proposed mechanisms is the need to change the STM application to be able to use it. Only the works proposed by Góes et al.~\cite{Goes:2012,Goes:2014} have this need, that is, the STM application needs to be created or modified to be able to benefit from the proposed optimizations. All other proposed mechanisms only change the STM runtime. Hence, it is unnecessary to change the STM applications.

\section{Limitations of existing proposals and future research directions}\label{sect:directions}

In this section, we will discuss some limitations and opportunities of research regarding thread and data mapping in Transactional Memory. 

The first observation is that no work is focusing on HTM or hybrid TM systems, only STM. Furthermore, all works used the \texttt{TinySTM}~\cite{TinySTM2} library as a base TM to implement the proposed mechanisms. Hence, it is not clear how much thread and data mapping can be helpful for other systems or STM implementations. The same observation can be made for distributed memory systems. All works presented in this survey are focused on STM running on shared memory systems.

Except for the work of Pasqualin et al.~\cite{Pasqualin:2020} there are no works that verified the relationship between aborts and thread mapping, that is, if they influence each other. In addition, the discussion presented in~\cite{Pasqualin:2020} only pointed out that there can be a relationship between them. However, it is necessarily an in-depth investigation, analyzing each application individually to draw strong conclusions about the topic. This leads us to another research opportunity: how thread and data mapping behave when used with techniques that focus on reducing the number of aborts, such as transactional schedulers~\cite{Hendler:2015,Sanzo:2017}. This relationship was also not studied in any related work. Similarly, there are no works that measured how much energy was saved by the optimized thread and data mapping using only the information provided by the STM runtime.

Not specific about thread and data mapping, but an issue that can influence experiments testing proposed mechanisms is the lack of new and updated benchmarks for TM systems. By far the most used benchmark suite for TM is \texttt{STAMP}~\cite{STAMP}. However, as pointed out in many works, \texttt{STAMP} is not ideal for testing online thread and data mapping since all threads have a similar behavior, executing the same transactional code, and the sharing behavior stays the same during execution time~\cite{Pasqualin:Bench,Castro:2012,Castro:2014,Goes:2014,Zhou:2016,Zhou:2018}. It would be interesting to change or add new applications to \texttt{STAMP} that present distinct sharing behaviors during execution or that have groups of threads doing different transactional jobs.

For online thread and data mapping mechanisms, it is necessary to extract information about the application during runtime with a low overhead. For this purpose, to our knowledge, only two works have been proposed~\cite{Castro:2011a,Schindewolf:2012}.

Finally, regarding machine learning (ML) techniques, only Castro et al.~\cite{Castro:2011,Castro:2012,Castro:2014} explored this approach. However, their mechanism relies on offline techniques, which require a previous execution of the application to collect the data for the training phase. Recent works on thread and data mapping for general applications successfully used ML techniques in an online context~\cite{Barrera:2020}. Hence, this online ML approach could be revisited in the context of transactional memory.

\section{Conclusions}\label{sect:conclusion}
Due to complex memory hierarchies and different latencies of memory accesses on multicore processors, locality of memory accesses is an essential issue for parallel application performance. Thread and data mapping are techniques to improve locality based on using information about the memory access behavior of a parallel application in order to optimize the location of threads and memory pages on NUMA systems. STM is an abstraction for thread synchronization that presents new challenges and opportunities for different types of thread and data mapping.

This paper surveyed proposals that solely use the information provided by an STM runtime to guide thread and data mapping. We found that an STM runtime can provide sufficient information to perform an efficient and effective thread mapping. However, our survey results indicate that only taking transactional information into consideration is insufficient to perform an efficient data mapping, since the STM runtime can only discover a fraction of the entire memory accessed by an application. Finally, we discussed research opportunities and directions where this research area could be expanded.
Some of the research opportunities for thread and data mapping in the TM domain, for instance, are on HTM, integration with transactional schedulers, energy efficiency, and machine learning, among others.

\bibliographystyle{plain}
\bibliography{references}

\end{document}